\documentclass[conference, 10pt]{IEEEtran}

\usepackage{cite}
\usepackage{url}
\usepackage{graphicx}
\usepackage{color}
\usepackage{placeins}
\usepackage{float}
\usepackage{tabularx,colortbl}
\usepackage{ifthen}
\usepackage{bm}
\usepackage{bbold}
\usepackage{amsmath}
\usepackage{mhchem}

\makeatletter

\newcounter{author}
\renewcommand{\author}[2][]{
   \stepcounter{author}
   \@namedef{author@\theauthor}{#2}
   \@namedef{authorlabel@\theauthor}{#1}
}

\newcounter{address}
\newcommand{\address}[2][]{
   \stepcounter{address}
   \@namedef{address@\theaddress}{#2}
   \@namedef{addresslabel@\theaddress}{#1}
}

\newcommand{\alsep}{and}

\def\newmaketitle{\par%
  \begingroup%
  \normalfont%
  \def\thefootnote{}
  \def\footnotemark{}
  \let\@makefnmark\relax
  \footnotesize
  \footnotesep 0.7\baselineskip
  \normalsize%
  \twocolumn[\thenewmaketitle\@IEEEaftertitletext]%
  \if@IEEEusingpubid
     \enlargethispage{-\@IEEEpubidpullup}%
  \fi
  \endgroup
  \setcounter{footnote}{0}\let\maketitle\relax\let\@maketitle\relax
  \gdef\@thanks{}%
  \let\thanks\relax}

\def\thenewmaketitle{
  \newpage
  \begin{center}%
    \vskip0.2em{\Huge\@IEEEcompsoconly{\sffamily}\@IEEEcompsocconfonly{\normalfont\normalsize\vskip 2\@IEEEnormalsizeunitybaselineskip
   \bfseries\large}\@title\par}\vskip1.0em\par%
    \vspace{1ex}
    \newcounter{c@author}
    \newcounter{c@tmp}
    \ifthenelse{\value{author}=2}{%
      \newcommand{\liand}{ and }}{%
      \newcommand{\liand}{, and }}
    \ifthenelse{\value{address}<2}{%
      \@nameuse{author@1}%
      \stepcounter{c@author}%
      \whiledo{\value{c@author}<\value{author}}{%
        \setcounter{c@tmp}{\value{author}}%
        \addtocounter{c@tmp}{-\value{c@author}}%
        \ifthenelse{\value{c@tmp}=1}{%
          \renewcommand{\alsep}{\liand}}{\renewcommand{\alsep}{, }}%
        \stepcounter{c@author}\alsep \@nameuse{author@\thec@author}}\\%
    }
    {
      \@nameuse{author@1}${}^{(\ref{\@nameuse{authorlabel@1}})}$%
      \stepcounter{c@author}%
      \whiledo{\value{c@author}<\value{author}}{%
      \setcounter{c@tmp}{\value{author}}%
      \addtocounter{c@tmp}{-\value{c@author}}%
      \ifthenelse{\value{c@tmp}=1}{%
        \renewcommand{\alsep}{\liand}}{\renewcommand{\alsep}{, }}%
      \stepcounter{c@author}\alsep \@nameuse{author@\thec@author}%
        ${}^{(\ref{\@nameuse{authorlabel@\thec@author}})}$%
      }
    }
    \vspace{0.2ex}

    \ifthenelse{\value{address}>0}{%
      \ifthenelse{\value{address}=1}{
        {\@nameuse{address@1}}
      }
      {
        \newcounter{c@address}

        \begin{center}
        \whiledo{\value{c@address}<\value{address}}
        {
          \refstepcounter{c@address}
            ${}^{(\thec@address)}$\,%
              \label{\@nameuse{addresslabel@\thec@address}}%
              \@nameuse{address@\thec@address}\\ %
        }
        \end{center}
      } 
    }
    {
      \relax
    }
  \end{center}
}

\makeatother

\title{Fast and Accurate Method for Doppler Averaging of Rydberg EIT Signals}

\author[org1]{Omar Nagib}
\author[org1]{Thad G. Walker}

\address[org1]{Department of physics, University of Wisconsin-Madison, USA, onagib@wisc.edu}

\begin{document}

\newmaketitle

\begin{abstract}
Modeling the effect of Doppler broadening due to the thermal atomic motion of Rydberg sensors typically relies on sampling a large velocity class, solving the equations of motion for each velocity, and then averaging the atomic density matrix over that velocity class. This process is inexact, slow, and presents a bottleneck in simulating Rydberg sensors. We present an approach that allows for fast, exact velocity averaging just by solving the equations of motion twice. We find the “propagator” that acts on the zero-velocity solution to generate the velocity-dependent atomic state for all velocities. By averaging this propagator over the Maxwell-Boltzmann velocity distribution of the atoms, we obtain an explicit, analytic formula that generates the averaged atomic state. This method is expected to save memory and time computational resources by one to several orders of magnitude, compared to traditional sampling approaches. 

\end{abstract}

\section{Background}

The use of Rydberg atoms as sensitive detectors of radio- and microwave- frequency radiation is a subject of intense current research. Such Rydberg electrometers typically consist of a room-temperature vapor of Rubidium or Cesium atoms, excited by two or three narrow-band lasers to a particular Rydberg state. The presence of electromagnetic fields causes resonant transitions between that Rydberg state and a neighboring one. The change of populations of the Rydberg states, due to an external electromagnetic field, is sensitively detected by a change in the transmission of one of the lasers, often using a quantum effect called Electromagnetically Induced Transparency (EIT) \cite{practicalEIT}. As an aid to experimental design, theoretically understanding and quantifying Rydberg detector performance is important. They are described accurately by a multi-state quantum mechanical model of the EIT process, where the state of the atoms is given by a density matrix \cite{EITtheoryandExp}.

Because these sensors operate at room temperature, the atoms are moving with different velocities under a Maxwell-Boltzmann distribution. This gives rise to Doppler broadening, which causes decoherence and degradation in the performance of a Rydberg sensor. Since it is a dominant source of error, it is essential to theoretically include it when modeling Rydberg sensors \cite{Raman_Ramsey}. Several approaches have been developed to tackle Doppler broadening, namely analytical \cite{Javan_2002,Javan_2003,EITtheory,EITtheoryandExp,Stationary}, phenomenological \cite{pheno}, and numerical approaches \cite{numerical,RydIQule, InverseSample}. While analytical approaches are illuminating, their domain of applicability is limited because the expressions are only valid under certain approximations and regimes, which is not true for all Rydberg sensors. Conversely, numerical approaches can handle more general cases, but at the cost of requiring the equations of motion to be solved for a large set of distinct velocities. This is followed by doing a Riemann sum to find the velocity-averaged density matrix. This process is inherently inexact, slow, and is complicated by the fact that EIT velocity dependencies often oscillate rapidly over a range of a few meters per second, requiring a fine velocity step to get sufficient accuracy. The computational overhead for averaging over atomic motion is typically several orders of magnitude \cite{RydIQule}. 

We present a general approach that does the velocity averaging exactly, given the solutions to a pair of eigenvalue problems.
Using these two solutions, we obtain an explicit, analytic formula that exactly computes the averaged atomic state, without the need for any sampling.

\section{Summary of the main result}

\begin{figure}[]
   \centering
    \includegraphics[width=0.9\columnwidth]{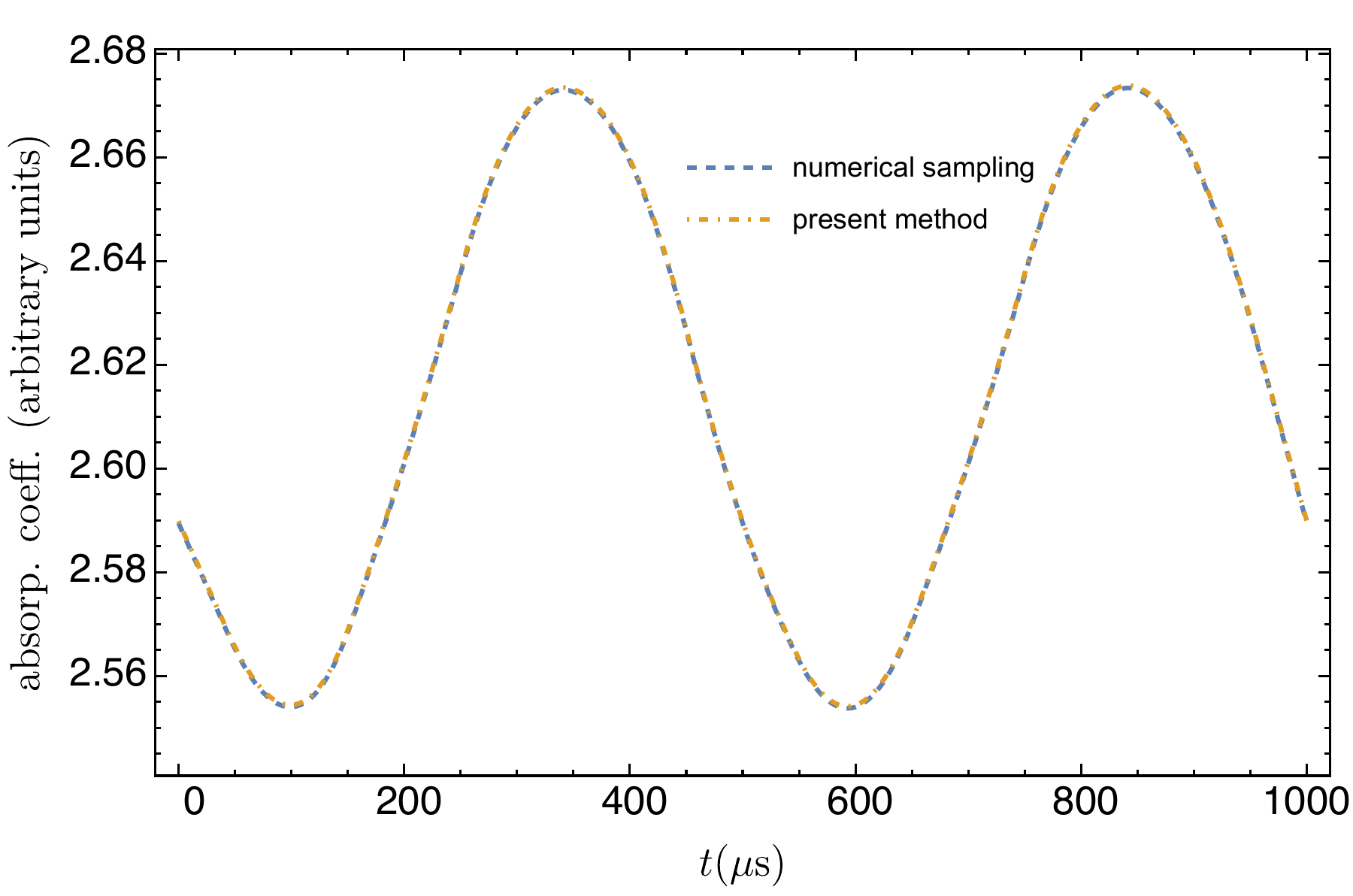}
   \caption{Absorption for a three-level EIT system of $\ce{^{87}Rb}$ versus time, using numerical sampling and the present method. The system parameters are $\Gamma= 2\pi \times 6 \ {\rm MHz}$, \ $v_{\rm th}=169.5 \ {\rm \mu m/\mu s}, \ k_1/2\pi= 1.28 \ {\rm \mu m} ^{-1},$ $k_2/2\pi=-1/1.248 \ {\rm \mu m}^{-1}$, and $\Omega_c/2\pi=\Omega_p/2\pi=f= 1 \ {\rm MHz}$. All detunings are set to zero.}
    \label{doppler_three_level_EIT}
\end{figure}

At steady state, the Rydberg sensor is described by the following linear equation:
\begin{equation}\label{eom}
(G_0+v G_v)\bm{\rho}_v=0,
\end{equation}
where $\bm{\rho}_v$ is the ``vectorized" atomic density matrix of an atom moving with a velocity $v$ (i.e., $\bm{\rho}_v$ is obtained from the usual density matrix $\rho_v$ by converting it into a column vector). $G_0$ includes the atomic structure, atom-laser interactions, and dissipation by spontaneous emission and collisions.  The primary effect of the atomic velocity is to produce Doppler shifts of the atomic resonance frequencies, quantified by the matrix $G_v$. For an $N$-level atomic system, $G_0$ and $G_v$ are $N^2 \times N^2$ matrices. For a given $v$, the solution of (\ref{eom}) corresponds to the null-space of $G_0+v G_v$. The velocity distribution is given by a Maxwell-Boltzmann distribution $P(v)=(2\pi v_{\rm th}^2)^{-1/2}\exp(-v^2/ 2v_{\rm th}^2)$,
where $v_{\rm th}$ is the thermal velocity of the atoms. Doppler broadening is calculated by averaging the velocity-dependent solution over that distribution, i.e.,
$\bar{\bm{\rho}}= \int P(v) \bm{\rho}_v dv.$
The following is the first key result that allows for an exact calculation of the Doppler averaged state. We show that the velocity-dependent solution $\bm{\rho}_v$ can be generated by acting with a “propagator” on the zero-velocity atomic state $\bm{\rho}_0$:
\begin{equation}\label{rhov}
\bm{\rho}_v=(\mathbb{1}+v G_0^-G_v)^{-1}\bm{\rho}_0,
\end{equation}
where $\bm{\rho}_0$ is the null-space of $G_0$. $G_0^-$ is a pseudo-inverse of $G_0$, constructed as follows. First, $G_0$ admits an eigenvalue decomposition in terms of its left and right eigenvectors and their eigenvalues. $G_0^-$ is then directly constructed by inverting $G_0$, excluding the zero-eigenvalue and zero-eigenvectors. To do the averaging, $G_0^- G_v$ is diagonalized to find its left and right eigenvectors, $\bm{r}_\lambda$ and $\bm{l}_\lambda$, and their eigenvalues $\lambda/v_{\rm th}$. Then $\bm{\rho}_v$ can be  decomposed as:
\begin{equation}
\bm{\rho}_v= \left( \sum_\lambda \frac{1}{1+\lambda v/v_{\rm th}} (\bm{r}_\lambda \otimes \bm{l}^{\rm T}_\lambda) \right) \bm{\rho}_0.
\end{equation}
The velocity dependence only appears in the scalar terms $1/(1+\lambda v/v_{\rm th})$, which can be integrated exactly over the velocity distribution. Carrying out the integration, we obtain an analytic expression for the exact, velocity-averaged density matrix:
\begin{equation}\label{rho_gauss_1D}
\bar{\bm{\rho}}=   \sum_\lambda \frac{\sqrt{\pi/2}}{\sqrt{-\lambda^2}} e^{-\frac{1}{2 \lambda^2}}\bigg(1+{\rm erf}\bigg[\dfrac{\sqrt{-\lambda^2}}{\sqrt{2}\lambda^2}\bigg]\bigg)  (\bm{r}_\lambda \otimes \bm{l}^{\rm T}_\lambda) \bm{\rho}_0.
\end{equation}
To summarize the averaging procedure:
\begin{enumerate}
    \item Diagonalize $G_0$ to find $\bm{\rho}_0$ and its eigenvectors and eigenvalues.
    \item Construct $G_0^-$ from $G_0$, and then find the eigenvectors and eigenvalues of $G_0^- G_v$. 
    \item The average is then given directly by (\ref{rho_gauss_1D}).
\end{enumerate}
As a non-trivial example, in Fig. \ref{doppler_three_level_EIT}, we compute the probe absorption versus time of a vapor cell of $\ce{^{87}Rb}$ atoms, for a three-level ladder EIT configuration. The ground and excited states are coupled by a control field $\Omega_c$, while the excited and the Rydberg state are coupled by a time-modulated field $\Omega_p \cos(2 \pi f t)$, where $\Omega_p$ and $f$ are the modulation amplitude and frequency, respectively. The two fields are counter-propagating. The wavenumbers for the one- and two- photon transitions are $k_1$ and $k_2$. The absorption is computed by the present approach and by numerically averaging over a uniform sample of $v$. Both methods agree, which confirms the validity of the present approach. The present approach, however, is faster by two orders of magnitude for this particular problem, and it does not require intensive memory resources. 
\section{Discussion and future directions}

Traditional numerical approaches need to solve the equations of motion repeatedly for a large sample of $v$, i.e., find the null-space of $G_0+vG_v$ for every $v$, and then approximate the average as a Riemann sum over that sample. The number of samples can be in the order of hundreds for Rydberg sensors with complicated EIT configuration \cite{RydIQule}. Instead, our approach reduces the problem to two diagonalizations, which can be done either numerically or analytically. Moreover, the averaging procedure itself is exact, even when the diagonalizations can only be obtained numerically. Therefore, our approach is expected to save computational resources by one to several orders of magnitude. The method is readily generalizable to efficiently model other fluctuations affecting the atomic state, e.g., fluctuations of the laser intensity and phase. Moreover, it can be extended to other physical systems beyond atomic systems. Two future research directions are worth mentioning. The first is extending the present approach when there is more than one random variable present in the problem. The second is investigating whether this approach can be generalized to the entire time evolution of the atomic state, not just its steady state.

\bibliographystyle{IEEEtran} 
\bibliography{doppler}

\end{document}